\documentclass[12pt]{article}

\usepackage{graphicx}

\usepackage{times}

\newenvironment{sciabstract}{%
\begin{quote} \bf}
{\end{quote}}

\newcounter{lastnote}

\title{Detection of Gamma-Ray Emission from the Vela Pulsar Wind Nebula with AGILE\\
\mbox{}\\
\footnotesize{\textsc{This manuscript has been accepted for publication in Science. This version has not undergone final editing. Please refer to the complete version of record at http://www.sciencemag.org/.  The manuscript may not be reproduced or used in any manner that does not fall within the fair use provisions of the Copyright Act without the prior, written permission of AAAS.}}}

\author
{A. Pellizzoni,$^{1\ast}$ A. Trois,$^{2}$ M. Tavani,$^{2,3,12,18}$ M. Pilia,$^{4,1}$ A. Giuliani,$^{5}$ G. Pucella,$^{15}$ \\
P. Esposito,$^{5,6}$ S. Sabatini,$^{2,12}$ G. Piano,$^{2,12}$ A. Argan,$^{2}$ G. Barbiellini,$^{7}$ A. Bulgarelli,$^{8}$ \\
M. Burgay,$^{1}$ P. Caraveo,$^{5}$ P. W. Cattaneo,$^{6}$ A. W. Chen,$^{5}$ V. Cocco,$^{2}$ T. Contessi,$^{5}$ \\
E. Costa,$^{2}$ F. D'Ammando,$^{2,3}$ E. Del Monte,$^{2}$ G. De Paris,$^{2}$ G. Di Cocco,$^{2}$ G. Di Persio,$^{2}$ \\
I. Donnarumma$^{2}$, Y. Evangelista,$^{2}$ M. Feroci,$^{2}$ A. Ferrari,$^{18}$ M. Fiorini,$^{5}$ F. Fuschino,$^{8}$ \\
M. Galli,$^{9}$ F. Gianotti,$^{8}$ A. Hotan,$^{10}$ C. Labanti,$^{8}$ I. Lapshov,$^{2}$ F. Lazzarotto,$^{2}$ \\
P. Lipari$^{11}$, F. Longo,$^{7}$ M. Marisaldi,$^{8}$ M. Mastropietro,$^{2}$ S. Mereghetti,$^{5}$  E. Moretti,$^{7}$ \\
A. Morselli,$^{12}$ L. Pacciani,$^{2}$ J. Palfreyman,$^{13}$ F. Perotti,$^{5}$  P. Picozza,$^{3,12}$ C. Pittori,$^{14}$\\
A. Possenti,$^{1}$ M. Prest,$^{4}$ M. Rapisarda,$^{15}$ A. Rappoldi,$^{6}$ E. Rossi,$^{8}$ A. Rubini,$^{2}$ \\
P. Santolamazza,$^{14}$ E. Scalise,$^{2}$ P. Soffitta,$^{2}$ E. Striani,$^{3,12}$ M. Trifoglio,$^{8}$ E. Vallazza,$^{7}$ \\
S. Vercellone,$^{16}$ F. Verrecchia,$^{14}$  V. Vittorini,$^{3}$ A. Zambra,$^{5}$ D. Zanello,$^{11}$ P. Giommi,$^{14}$ \\
S. Colafrancesco,$^{14}$ A. Antonelli,$^{14}$ L. Salotti,$^{17}$ N. D'Amico,$^{1,19}$ G. F. Bignami$^{20}$ \\
\\
\footnotesize{$^1$ INAF - Osservatorio Astronomico di Cagliari, loc. Poggio dei Pini, strada 54, I-09012, Capoterra (CA), Italy}\\
\footnotesize{$^2$ INAF--IASF Roma, via del Fosso del Cavaliere 100, I-00133 Roma, Italy}\\
\footnotesize{$^3$ Dipartimento di Fisica, Universit\`a Tor Vergata, via della Ricerca Scientifica 1, I-00133 Roma, Italy}\\
\footnotesize{$^4$ Dipartimento di Fisica, Universit\`a dell'Insubria, via Valleggio 11, I-22100, Como, Italy}\\
\footnotesize{$^5$ INAF--IASF Milano, via E. Bassini 15, I-20133 Milano, Italy}\\
\footnotesize{$^6$ INFN--Pavia, via A. Bassi 6, I-27100 Pavia, Italy}\\
\footnotesize{$^7$ INFN--Trieste, Padriciano 99, I-34012 Trieste, Italy}\\
\footnotesize{$^8$ INAF--IASF Bologna, via P. Gobetti 101, I-40129 Bologna, Italy}\\
\footnotesize{$^9$ ENEA Bologna, via don G. Fiammelli 2, I-40128 Bologna, Italy}\\
\footnotesize{$^{10}$ Curtin University of Technology, 78 Murray Street, Perth, WA 6000, Australia}\\
\footnotesize{$^{11}$ INFN--Roma La Sapienza, p.le A. Moro 2, I-00185 Roma, Italy}\\
\footnotesize{$^{12}$ INFN--Roma Tor Vergata, via della Ricerca Scientifica 1, I-00133 Roma, Italy}\\
\footnotesize{$^{13}$ School of Mathematics and Physics, University of Tasmania, Hobart, TAS 7001, Australia}\\
\footnotesize{$^{14}$ ASI Science Data Center, ESRIN, I-00044 Frascati (Roma), Italy}\\
\footnotesize{$^{15}$ ENEA Frascati, via E. Fermi 45, I-00044 Frascati (Roma), Italy}\\
\footnotesize{$^{16}$ INAF--IASF Palermo, via U. La Malfa 153, I-90146 Palermo, Italy}\\
\footnotesize{$^{17}$ ASI - Agenzia Spaziale Italiana, viale Liegi 26, I-00198 Roma, Italy}\\
\footnotesize{$^{18}$ Consorzio Interuniversitario per la Fisica Spaziale, viale Settimio Severo 63, I-10133 Torino, Italy}\\
\footnotesize{$^{19}$ Dipartimento di Fisica, Universit\`a di Cagliari, Cittadella Universitaria, I-09042 Monserrato (CA), Italy}\\
\footnotesize{$^{20}$ Istituto Universitario di Studi Superiori, viale Lungo Ticino Sforza 56, I-27100 Pavia, Italy}\\
\footnotesize{$^\ast$To whom correspondence should be addressed. E-mail:  apellizz@ca.astro.it}
}

\date{}

\begin{document} 


\baselineskip24pt


\maketitle


\begin{sciabstract}

Pulsars are known to power winds of relativistic particles that can produce
bright nebulae by interacting with the surrounding medium.
These pulsar wind nebulae (PWNe) are observed in the radio, optical,
x-rays and, in some cases, also at TeV energies,
but the lack of information in the gamma-ray band prevents from
drawing a comprehensive multiwavelength picture of their phenomenology
and emission mechanisms.
Using data from the AGILE satellite, we detected the Vela
pulsar wind nebula in the energy range from 100 MeV to 3 GeV. 
This result constrains the particle population responsible for the GeV emission, probing
multivavelength PWN
models, and establishes a class of
gamma-ray emitters that could account for a fraction of the
unidentified Galactic gamma-ray sources.

\end{sciabstract}



The Vela supernova remnant (SNR) is the nearest SNR ($d\simeq290$ pc) containing a bright pulsar, PSR B0833--45, which has
a characteristic age of 11 kyr, and a spin-down
luminosity of $7\times10^{36}$ erg s$^{-1}$ (\emph{\ref{taylor93}, \ref{dodson03}}).
This SNR extends over a diameter of $\sim$8$^{\circ}$, and is known from
early radio observations to embrace a number of regions of non-thermal emission (\emph{\ref{rishbeth58}}) including Vela X,
a $100'$-diameter, flat-spectrum radio component near the
center of the SNR. Vela X, separated by $\sim$$40'$ from PSR B0833--45, is generally interpreted as the pulsar's radio synchrotron nebula (\emph{\ref{weiler80}, \ref{dwarakanath}}).
A diffuse emission
feature ($\sim$1$^{\circ}$ long) coincident with the centre of Vela X was detected in x-rays (0.6--7.0 keV) by the R\"ontgen
(\emph{\ref{markwardt95}}) and ASCA (\emph{\ref{markwardt97}}) satellites.
It was first suggested that this feature, which is
 closely aligned with a filament detected at radio wavelengths,
corresponds to the outflow jet from the pulsar's pole (\emph{\ref{frail97}}). More recently, observations with Chandra (\emph{\ref{helfand01}}) clearly unveiled the torus-like morphology of the compact x-ray nebula surrounding the pulsar and
indicated that the centre of Vela X  lies along the extension
of the pulsar equator, although bending to the southwest.

The detection of very high-energy (VHE; 0.5--70 TeV) gamma-rays from the Vela X region was claimed by HESS (\emph{\ref{aharonian06}}) and confirmed by CANGAROO (\emph{\ref{enomoto06}}).
The strong VHE-source, HESS J0835--455 (luminosity of $\sim$10$^{33}$ erg s$^{-1}$ at energies above 0.55 TeV),
coincides with the region of hard x-ray emission seen by the R\"ontgen Satellite.
The best-fit VHE emission centroid ($\rm Ra.=08^{\rm h}35^{\rm m}1^{\rm s}$, $\rm Decl. =-45^{\circ}34'40''$)
is $\sim$0.5$^{\circ}$ from the pulsar position
 and the VHE emission has an extension of $\sim$5$\times$4 parsec$^{2}$.
The detection of Vela X at TeV energies demonstrated that this
source emits non-thermal radiation, in agreement with the hypothesis that it corresponds
to the pulsar wind nebula, displaced to the south by the unequal pressure of the reverse
shock from the SNR (\emph{\ref{blondin01}}).

The multiwavelength spectrum of the center of Vela X can be modeled as synchrotron radiation
from energetic electrons within the cocoon (radio and x-rays) and
inverse Compton emission from the scattering (by the same
electron population) of the cosmic microwave background
radiation (CMBR), the Galactic far-infrared radiation (FIR) produced by reradiation of dust grains, and the local starlight (\emph{\ref{dejager07}, \ref{lamassa08}, \ref{dejager08}}).
Alternatively, a hadronic model can be invoked for the gamma-ray
emission from the Vela X cocoon, where the emission is the result of the decay of
neutral pions produced in proton-proton collisions (\emph{\ref{horns06}}).
Observations in the MeV-GeV band (HE) are crucial to distinguish between leptonic and hadronic models as well as to identify specific particle populations and spectra. \\
\indent The Vela region was recently observed from 30 MeV to 50 GeV 
by the AGILE (\emph{\ref{pellizzoni09}}) and Fermi (\emph{\ref{abdo09}})
gamma-ray satellites. The Vela pulsar is the brightest persistent source of the GeV sky
 and, due to the limited angular resolution of the current-generation gamma-ray instruments, its 
gamma-ray pulsed  emission dominates the surrounding region up to a
radius of about 5$^{\circ}$,
preventing the effective identification 
of weaker nearby sources.\\
\indent The AGILE satellite (\emph{\ref{tavani09}}) observed the Vela pulsar for $\sim$180 days (within 60$^{\circ}$ from the center of instrument's
field of view) from 2007 July (54294.5 MJD) to 2009 September (55077.7 MJD).
To obtain precise radio ephemeris and model the Vela pulsar timing noise for the entire AGILE data span, we made use of observations with the Mount Pleasant radio telescope.
The Vela pulsar timing analysis provided a total of $\sim$40000 pulsed counts with energies between 30 MeV and 50 GeV; the difference between the radio and gamma-ray ephemeris was $<$$10^{-11}$ s.
Gamma-ray pulsed counts are concentrated within the phase interval 0.05--0.65 (where 0 is the phase
corresponding to the main radio peak).
We verified that no pulsed gamma-ray emission is detected outside this interval, as reported by previous observations with EGRET (\emph{\ref{kanbach94}, \ref{thompson04}}), AGILE (\emph{\ref{pellizzoni09}}) and Fermi (\emph{\ref{abdo09}}).

With the aim of performing a sensitive search for close faint sources excluding the bright emission from the Vela pulsar, we discarded the time intervals corresponding to the phase interval 0.05--0.65.
The analysis of the resulting off-pulse images (taking only events
corresponding to 0.65--1.05 pulsar phase interval, for a total of $\sim$ 14000 events),
unveiled few gamma-ray sources, none of which coincides with the Vela
pulsar.
A maximum likelihood analysis (\emph{\ref{tavani09}}), performed on the $E>100$ MeV dataset within a region of
5$^{\circ}$ around the pulsar position, revealed two sources at better than 3$\sigma$ confidence (see Figure 1):
AGL J0848--4242 (at Galactic coordinates,
$l=263.11^{\circ}$, $b=0.65^{\circ}$, 68\% confidence error circle radius $\sim$0.25$^{\circ}$) and AGL J0834--4539
(at $l=263.88^{\circ}$, $b=-3.17^{\circ}$, e.c. radius $\sim$0.20$^{\circ}$).
A gamma-ray source coincident with the EGRET source 3EG J0841--4356 (\emph{\ref{hartman99}}) was also detected with lower significance,
and the Vela Junior (RX J0852.0--4622) SNR (\emph{\ref{aschenbach98}, \ref{aharonian05}}) is also possibly contributing to  a counts excess in the Galactic Plane
around $l$$\sim$265.6$^{\circ}$.

The brightest gamma-ray source, AGL J0834--4539
($\sim$5.9$\sigma$ significance, $\sim$264 counts, $F\gamma=(35\pm7)\times10^{-8}$ ph cm$^{-2}$ s$^{-1}$ at $E>100$ MeV), is 
located $\sim$0.5$^{\circ}$ southwest from the Vela pulsar position (outside the 95\% source position confidence contour) and has a spatial extent of $\sim$1.5$\times$1 square degrees. Its shape is asymmetric and incompatible with the AGILE point-spread function. Therefore, possible residual emission from the pulsar (in principle associated to undetected weak peaks in the off-pulse interval of the light curve) cannot substantially contribute to this diffuse feature.
No relevant systematic errors on positions, fluxes and spectra (mostly due to uncertainties on the Galactic gamma-ray diffuse emission model) affect AGILE sources detected around 5$\sigma$ level.
AGL J0834--4539 is positionally concident with HESS J0835--455, the TeV source that is identified with the Vela X nebula, and has a similar brightness profile to it (Figure 1). This implies that AGL J0834--4539 is associated with the pulsar's PWN.\\
\indent Based on the available count statistics, we performed a first estimate of the spectrum by sampling the flux
in the three energy bands (0.1--0.5 GeV, 0.5--1 GeV, 1--3 GeV; see Figure 2) where
the source is clearly detected.
A power law fit yields a photon index
$\alpha=-1.67\pm0.25$.  
The AGILE spectral points are a factor $\sim$2 below the previous EGRET upper limits (\emph{\ref{dejager96}}) and  well above the extrapolation of
HESS $\nu F_{\nu}$ spectrum to lower energies.
The PWN gamma ray luminosity in the 0.1--10 GeV band, for a distance of $\sim$290 pc (\emph{\ref{dodson03}}, {\emph{\ref{caraveo01}}), is 
4$^{+4}_{-2}\times10^{33}$ erg s$^{-1}$
 corresponding to $\sim$10$^{-3}$ $\dot{E}_{rot}$ (where $\dot{E}_{rot}$ is the spin-down luminosity of the pulsar).
Such a luminosity is slightly higher than at VHE energies
($9.9\times10^{32}$ erg s$^{-1}$). \\
\indent In the frame of leptonic models, the AGILE measurements are not consistent with a simple multiwavelength spectral energy distribution
involving a single electron population.
The AGILE spectral points are one order of magnitude above the fluxes expected from the electron population  simultaneously fitting synchrotron x-ray emission (peaking at $\sim$1 keV)
and inverse-Compton (IC) TeV emission (\emph{\ref{aharonian06}, \ref{lamassa08}}).\\
\indent Additional electron populations should be invoked to explain the observed GeV
fluxes. This is not surprising in view of the complex morphology of the PWN
seen in radio and x-rays, where different sites and features of non-thermal emission are present:
the anisotropic pulsar wind and non-homogeneous SNR reverse shock pressure produce different particle populations within the shocked wind.
In particular, assuming the same magnetic field 
(5$\mu$G) reproducing the TeV spectral break, the radio synchrotron emitting electrons observed in the Vela X structure (\emph{\ref{alvarez01}}) may be responsible for the inverse Compton (IC) bump in the GeV band arising from 
scattering on CMBR and Galactic and starlight photon fields, as noticeably predicted  by de Jager et al. (\emph{\ref{dejager08}}, \emph{\ref{dejager07}}).
In fact, the position where AGILE sees the maximum brigthness ($\rm Ra.=08^{\rm h}35^{\rm m}$, $\rm Decl. =-45^{\circ}44'$)
is also roughly where the 8.4 GHz radio emission is brightest (\emph{\ref{hales04}}). AGILE data are compatible with the IC parameters modelled by de Jager et al. \emph(\ref{dejager08})
 (electron spectral index 1.78
and maximum energy $\sim$20 GeV), although our measurements could suggest a higher contribution from IC photon seeds.
In particular, assuming a starlight energy density of 1.4 eV cm$^{-3}$ and a mean temperature of $\sim$2300 K (\emph{\ref{giulianiweb}}),
we obtain a good description of the AGILE data (Figure 2).\\
\indent The AGILE measurements would be incompatible with the scenario of nucleonic gamma-ray production
in the Vela TeV nebula in the frame of a single primary electron population. These models predict very faint GeV
emission ($<$10$^{30}$ erg s$^{-1}$) even when including synchrotron and inverse Compton emission from primary
and secondary electrons produced by the inelastic nuclear scattering (\emph{\ref{horns06}}).
On the other hand, the proposed additional electron component scenario described above leaves room for
uncorrelated GeV-TeV emission, although the comprehensive multiwavelength two-component leptonic model
(providing strong IC emission on a relatively dense photon field) 
seems to disfavour dominant nucleonic gamma-ray production.
In fact, it has been found that the thermal particle density at the
head of the cocoon, where bright VHE gamma-ray emission was found, is a factor of 6 lower
than that required by hadronic models (\emph{\ref{lamassa08}}).\\
\indent The radio emitting region mentioned above appears to be larger ($\sim$2$\times$3 square degrees) than the AGILE nebula, possibly indicating that IC cooling in the GeV domain is important,
but we notice that the actual physical 
size of the GeV nebula could be larger than that presently resolved with 
the available photon statistics, due to the strong Galactic gamma-ray 
emission affecting MeV-GeV energy bands.
Instead, the AGILE nebula is similar in shape to the HESS nebula, possibly
 suggesting that the core of HE and VHE emission is produced in
the same projected region of Vela X, even if different electron populations are
involved.
Indeed, different spots of bright radio emission (\emph{\ref{hales04}}), possibly associated to electrons injected at different stages of pulsar evolution, are embedded within the poorly resolved HE and VHE emission regions.\\
\indent High-energy emitting PWNe are thought to be a common phenomenon associated
to young and energetic pulsars ({\emph{\ref{mattana09}}) because
their IC emission arises mostly from scattering on CMBR and starlight fields with no
special environmental requirements.
On the other hand, PWNe are expected to be much weaker than pulsed emission from the
associated neutron star, especially in the GeV domain where most of the pulsar's spin-down
energy is funnelled.

Indeed, in spite of a PWN gamma ray yield of $L_{\gamma}^{PWN}\sim$10$^{-3}\times\dot{E}_{rot}$,
to be compared with  the typical gamma-ray pulsed luminosity of $L_{\gamma}^{pulsed}\sim(10^{-2}$--0.1)$\times\dot{E}_{rot}$, our AGILE observation shows that 10 ky old PWNe can match the
sensitivities of current GeV instruments.

Because the gamma-ray luminosity of the PWNe is only a small
fraction of the beamed emission from the neutron star, the PWN component is difficult to 
identify in weaker gamma-ray pulsars, although it could account for a substantial part of the observed off-pulse flux.
However, if the beamed emission does not intersect the line of sight to the
observer, the PWN component, unhindered by the stronger pulsed emission, could be
detectable.
Energetic pulsars (e.g. $\dot{E}_{rot}\sim10^{37}$ erg s$^{-1}$) can power PWNe with 
gamma-ray luminosities
matching the flux ($\sim$10$^{-8}$--10$^{-7}$ ph cm$^{-2}$ s$^{-1}$; $E>100$ MeV) of a class of
  unidentified EGRET sources (\emph{\ref{hartman99}}), as well as a subset of the ones detected by AGILE and Fermi (\emph{\ref{pittori09}, \ref{abdo09cat}}), when
placed within few kiloparsecs.
The roughly isotropic emission from such undisturbed PWNe would not
yield pulsations and, as a class, they could contribute to the
population of Galactic unidentified sources still awaiting multiwavelength association (\emph{\ref{pellizzoni04}}, \emph{\ref{dejager07}}).

\begin{quote}
{\bf References and Notes}
\begin{enumerate}
\item \label{taylor93} 	J. H. Taylor, R. N. Manchester, A. G. Lyne, {\it Astrophys J. Supplement Series}  {\bf 88}, 529 (1993).
\item \label{dodson03}	R. Dodson, D. Legge, J. E. Reynolds, P. M. McCulloch, {\it Astrophys J.}  {\bf 596}, 1137 (2003).
\item \label{rishbeth58} H. Rishbeth, {\it Australian J. of Physics} {\bf 11}, 550 (1958).
\item \label{weiler80} K. W. Weiler, N. Panagia, {\it Astron. Astrophys.}  {\bf 90}, 269 (1980).
\item \label{dwarakanath} K. S. Dwarakanath,  {\it J. of Astrophys. Astron.}  {\bf 12}, 199 (1991).
\item \label{markwardt95} C. B. Markwardt, H. B. Oegelman, {\it Nature}  {\bf 375}, 40 (1995).
\item \label{markwardt97} C. B. Markwardt, H. B. Oegelman, {\it Astrophys J.}  {\bf 480}, 13 (1997).
\item \label{frail97} D. A. Frail, M. F. Bietenholz, C. B. Markwardt, H. Oegelman,  {\it Astrophys J.}  {\bf 475}, 224 (1997).
\item \label{helfand01} D. J. Helfand, E. V. Gotthelf, J. P. Halpern, 	  {\it Astrophys J.}  {\bf 556}, 380 (2001).
\item \label{aharonian06} F. Aharonian \emph{et al.}, {\it Astron. Astrophys.}  {\bf 448}, L43 (2006).
\item \label{enomoto06} R. Enomoto \emph{et al.}, {\it Astrophys J.}  {\bf 638}, 397 (2006).
\item \label{blondin01} J. M. Blondin, R. A. Chevalier, D. M. Frierson, {\it Astrophys J.}  {\bf 563}, 806 (2001).
\item \label{dejager07} O. C. de Jager, {\it Astrophys J.}  {\bf 658}, 1177 (2007).
\item \label{lamassa08}  S. LaMassa, P. O. Slane, O. C. de Jager, {\it Astrophys J.}  {\bf 689}, L121 (2008).
\item \label{dejager08} O. C. de Jager, P. O. Slane, S. LaMassa, {\it Astrophys J.}  {\bf 689}, L125 (2008).
\item \label{horns06} D. Horns, F. Aharonian, A. Santangelo, A. I. D. Hoffmann, C. Masterson, {\it Astron. Astrophys.}  {\bf 451}, L51 (2006).
\item \label{pellizzoni09} A. Pellizzoni \emph{et al.}, {\it Astrophys J.}  {\bf 691}, 1618 (2009).
\item \label{abdo09} A. Abdo \emph{et al.}, {\it Astrophys J.}  {\bf 696}, 1084 (2009).
\item \label{tavani09} M. Tavani  \emph{et al.}, {\it Astron. Astrophys.}  {\bf 502}, 995 (2009).
\item \label{kanbach94} G. Kanbach \emph{et al.}, {\it Astron. Astrophys.}  {\bf 289}, 855 (1994).
\item \label{thompson04} D. J. Thompson, {\it Cosmic Gamma-Ray Sources}, (Astrophysics and Space Science Library 304 ed. K. S. Cheng \& G. E. Romero (Dordrecht: Kluwer)), 149.
\item \label{hartman99} R. C. D. L. Hartman \emph{et al.}, {\it Astrophys J. Supplement Series} {\bf 123}, 79 (1999).
\item \label{aschenbach98} B. Aschenbach, {\it Nature}  {\bf 396}, 141 (1998).
\item \label{aharonian05} F. Aharonian \emph{et al.}, {\it Astron. Astrophys.}  {\bf 437}, L7 (2005).
\item \label{dejager96} O. C. de Jager, A. K. Harding, P. Sreekumar,  M. Strickman, {\it Astron. Astrophys. Supplement}  {\bf 120}, 441 (1996).
\item \label{caraveo01} P. A. Caraveo, A. De Luca, R. P. Mignani, G. F. Bignami, {\it Astrophys J.}  {\bf 561}, 930 (2001).
\item \label{alvarez01} H. Alvarez, J. Aparici, J. May, P. Reich, {\it Astron. Astrophys.}  {\bf 372}, 636 (2001).
\item \label{hales04} A. S. Hales \emph{et al.}, {\it Astrophys J.}  {\bf 613}, 977 (2004).
\item \label{giulianiweb} Stellar radiation field peak is generally assumed at $E$$\sim$1 eV well fitting the gamma-ray Galactic diffuse emission; see e.g.\\ http://www.iasf-milano.inaf.it/$\sim$giuliani/public/thesis/node10.html
\item \label{mattana09} F. Mattana \emph{et al.}, {\it Astrophys J.}  {\bf 694}, 12 (2009).
\item \label{pittori09} C. Pittori \emph{et al.}, {\it Astron. Astrophys.} {\bf 506}, 1563 (2009).
\item \label{abdo09cat} A. Abdo \emph{et al.}, {\it Astrophys J. Supplement Series}  {\bf 183}, 46 (2009).
\item \label{pellizzoni04} A. Pellizzoni \emph{et al.}, {\it AIP Conference Proceeding} (2nd International Symposium on High Energy Gamma-Ray Astronomy, Heidelberg) {\bf 745}, 371 (2005).
\item The AGILE mission is funded by the Italian Space Agency with scientific and programmatic
participation by the Italian Institute of Astrophysics and the Italian Institute of Nuclear Physics.
We acknowledge the anonymous referees for their useful comments on the paper draft.
\end{enumerate}
\end{quote}

\begin{figure}
\centering
\includegraphics[angle=00,width=10.5cm]{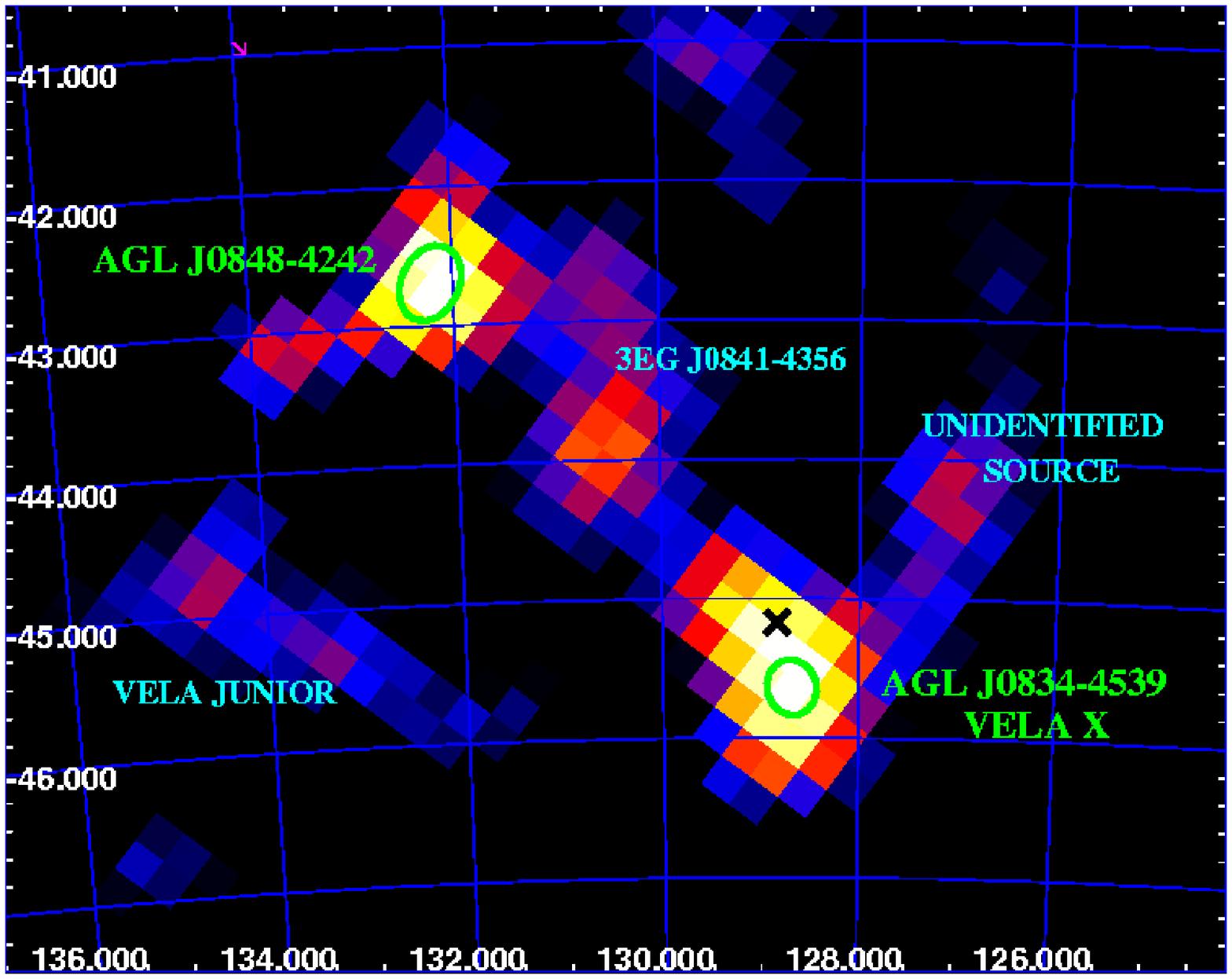}
\includegraphics[angle=00,width=10.5cm]{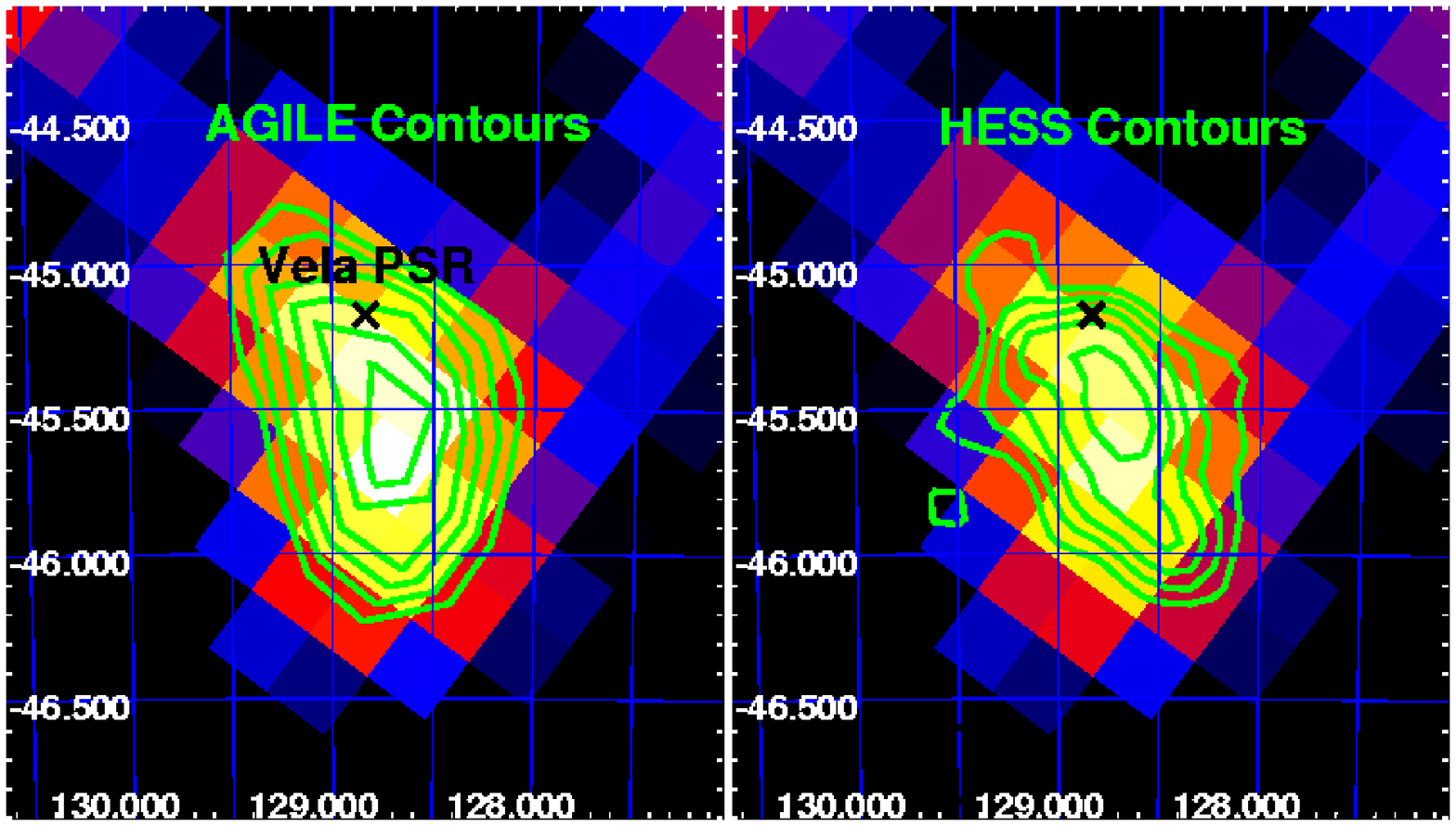}
\includegraphics[angle=00,width=10.5cm]{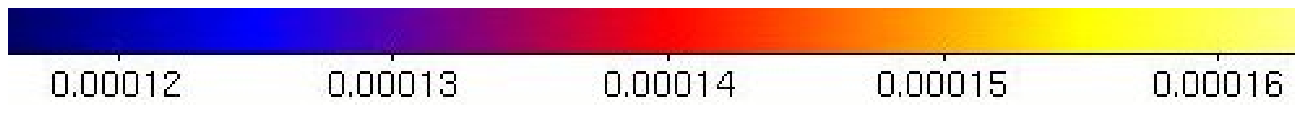}
\caption{ {\bf{$Top$}}: Gaussian-smoothed AGILE intensity map (ph cm$^{-2}$ s$^{-1}$ sr$^{-1}$ with 0.25$\times$0.25 square degrees pixel size) at $E>400$ MeV around the Vela pulsar, including only off-pulse events (i.e. discarding events with pulsar phase corresponding to Vela pulsed emission); the neutron star position is marked with the black cross, green circles are the 68\% confidence contours for the position of AGL J0848-4242 and AGL J0834-4539 [Vela X]. The AGILE $E>400$ MeV energy band is well suited for gamma-ray imaging and provides a good compromise between the instrument effective area ($\sim$400 cm$^{2}$; $\sim$100 counts from AGL J0834--4539) and point spread function ($\sim$1$^{\circ}$, 68\% containment radius), both parameters decreasing with energy. {\bf{$Bottom$}}: the gamma-ray diffuse source AGL J0834--4539. AGILE contours ($bottom-left$) are in the range (1.4--$1.6)\times10^{-4}$ ph cm$^{-2}$ s$^{-1}$ sr$^{-1}$, with step 4$\times$10$^{-6}$. HESS contours ($bottom-right$) are taken from (\emph{\ref{aharonian06}}).}
\end{figure}

\begin{figure}
\centering
\includegraphics[angle=00,height=12.0cm,width=15.0cm]{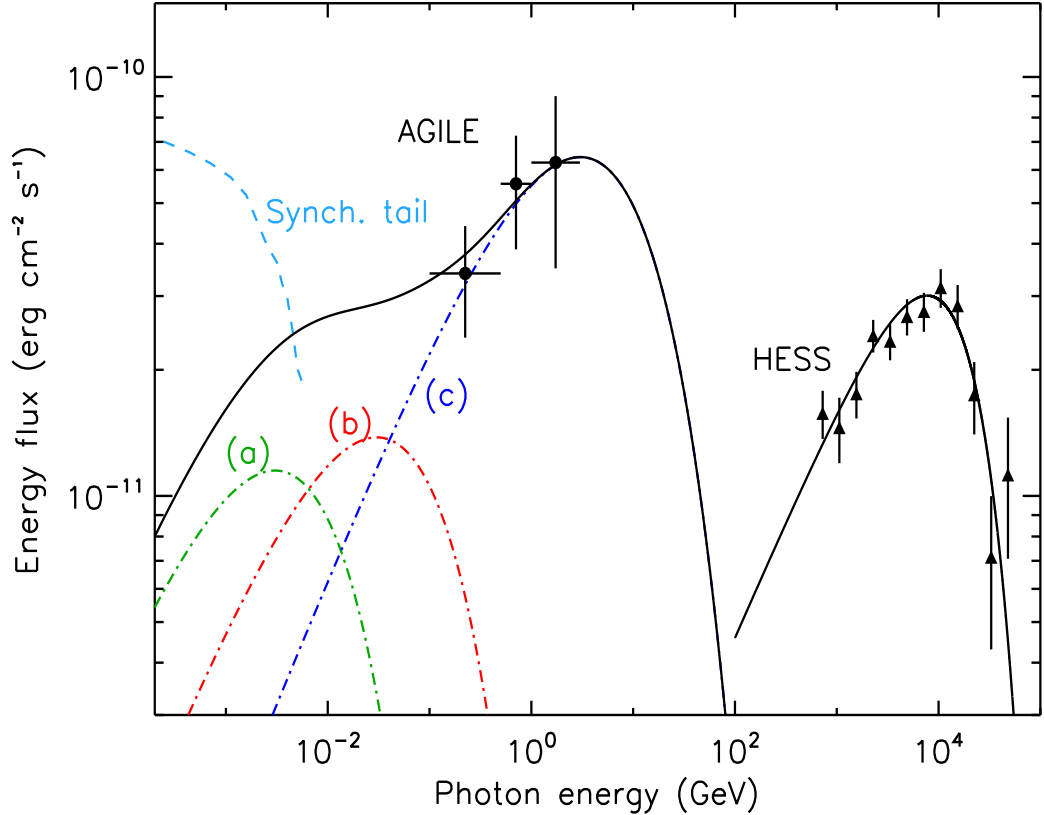}
\caption{Gamma-ray high- and very-high energy spectral  distribution ($\nu F_{\nu}$) of the Vela X PWN. HESS data fit an IC process (scattering on CMBR) related to electron power law index 2.0 with a break at 67 TeV and a total energy content of $2.2\times10^{45}$ erg (\emph{\ref{aharonian06}}). AGILE data are compatible with IC emission from the additional electron component well reproducing the observed total radio spectrum ($E_{\rm tot}=4\times10^{48}$ erg) assuming the same $\sim5$ $\mu$G field strength as required by the TeV spectral break. Unlike the TeV IC emission, GeV IC scattering is within Thomson limit. Thus, in addition to the CMBR component (photon density $n_{\rm ph}=0.25$ eV cm$^{-3}$, photon energy $E_{\rm ph}=10^{-3}$ eV), also FIR ($n_{\rm ph}=0.3$ eV cm$^{-3}$, $E_{\rm ph}=10^{-2}$ eV) and starlight ($n_{\rm ph}=1.4$ eV cm$^{-3}$, $E_{\rm ph}=1$ eV) photon fields can significantly contribute to the high-energy IC counterpart of the radio spectrum fitting AGILE data ($dot-dashed$ lines: CMBR $(a)$, FIR $(b)$, starlight $(c)$; $thick$ line: total IC spectrum).}
\end{figure}

\end{document}